\documentclass[11pt,twoside]{article}
\usepackage{asp2004}
\usepackage{psfig}
\usepackage{epsf}
\usepackage{natbib}
\usepackage{graphics}
\usepackage{lscape}
\def\edcomment#1{\iffalse\marginpar{\raggedright\sl#1\/}\else\relax\fi}
\reversemarginpar

\begin{document}
\title{Rotation and Mixing in Massive Stars: Principles and Uncertainties}
\author{Georges Meynet and Andr\'e Maeder}
\affil{Geneva Observatory, CH--1290 Sauverny, Switzerland}

\begin{abstract}
The main instabilities induced by rotation in stellar interiors are described.
We derive from first principles the general equation describing the transport of the
angular momentum.
The case of the transport of the chemical species is also discussed.
As long as the mass loss rates are not too important, meridional currents,
by advecting angular momentum from the inner regions to the outer layers, accelerate
the  stellar surface during the Main Sequence phase.
A 9 M$_\odot$ stellar model at solar metallicity with an equatorial velocity
at the beginning of the core H--burning phase equal to 340 km s$^{-1}$
reaches the break--up limit during
the MS phase. The model with an initial velocity of 
$290$ km s$^{-1}$ approaches this limit without reaching it.
The models with $290$ km s$^{-1}$ and 340 km s$^{-1}$
predict enhancements of the N/C ratio
at the end of the MS phase equal to 2.8 and 3.2 times the initial value
respectively.
\end{abstract}
\thispagestyle{plain}

\section{The main instabilities in a rotating star}

Recent discussions of the various instabilities induced by rotation may be found
in \citet{MMAA}, \citet{Hel00}, and \citet{Ta04}.
Among the most important instabilities are the secular shear instability and the meridional circulation.
These instabilities drive the transport of the chemical species and of the angular momentum.
Let us briefly recall the physical principles underlying these two instabilities.

In a rotating star
local radiative equilibrium cannot be achieved (\citeauthor{vZ24} \citeyear{vZ24}; 
\citeauthor{Ed25} \citeyear{Ed25}; \citeauthor{Vo25} \citeyear{Vo25}). 
As a result some parts of the star are heated while others are cooled. The
buoyancy forces then drive 
a large scale motion, called the meridional circulation.
Meridional circulation, contraction/exp\-an\-sion of the stellar layers, and convection
create gradients of the angular velocity inside the star. These
gradients  produce instabilities known as shear
instabilities. The physical reason for this instability lies in the
fact that the minimum energy state of a differentially rotating fluid is solid 
body rotation.
The star will tend to approach this state by homogenizing
the angular velocity by turbulent mixing.

In a radiative zone, the vertical stable density stratification counteracts both the shear and meridional instability.
In that respect the $\mu$--gradients play a key role as a stabilizing agent. These gradients may even, depending
on the physics involved in the model, completely inhibit the mixing \citep{Me97}. 
There are different methods in the literature for accounting for the effects of the $\mu$--gradients
on the mixing. Some authors choose a parametric approach consisting in multiplying the $\mu$--gradient by
a free parameter, $f_{\mu}$, smaller than one in order to weaken the stabilizing effect of the 
$\mu$--gradient. The value of $f_{\mu}$ is chosen in order to enable the stellar models
to reproduce the observed surface enrichments \citep{Hel00}. The very small values
chosen (of the order of $0.05$)  well illustrates the strong inhibiting effects of the $\mu$--gradients.
Other methods devised by \citet{Ma97}, \citet{Ta97} take into account of the effects of the strong horizontal turbulence (see below) and of the
heat exchanges which can occur during the mixing process. This reduces the vertical drawback force and makes
 easier the development of the instability.
In that case, the energy available 
in the shear can always be used for performing some mixing. Expressions for the shear
diffusion coefficients can then be deduced without the need of artificially reducing the inhibiting effect of
the $\mu$--gradients.

In the horizontal direction, i.e.
on a isobar, in contrast to what happens
in the vertical direction, the shear instability is not inhibited by a stable
density stratification and the turbulence can develop without difficulty.
Thus as long as horizontal gradients are continuously built up by e.g. meridional circulation,
a strong horizontal turbulence develops in the star \citep{Za92}.
This strong horizontal turbulence will also erode very  efficiently
any horizontal gradient of $\Omega$. Thus 
the star   
can always be considered 
to be in a state of shellular rotation, characterized by constant values of $\Omega$ on isobars. 
Starting from this a priori, but reasonable hypothesis, \citet{Za92} proposed
a consistent theory of the interaction between the shear instability and the meridional circulation.
The theory can be kept one dimensional thanks to the hypothesis of strong horizontal turbulence.

\section{Equations for the transport of the angular momentum}

\begin{figure}[!t]
\plotfiddle{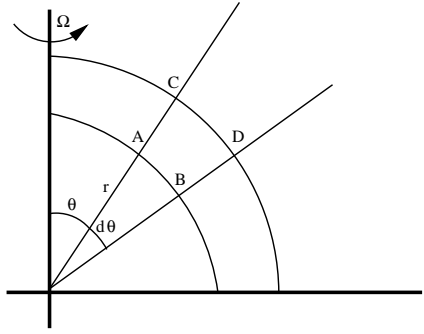}{5cm}{0}{30}{30}{-70}{0}
\caption{The momentum of the viscosity forces acting on the element ABCD is derived in the text and
the general form of the equation describing the change with time of the angular momentum of this element
is deduced. The star rotates around the vertical axis with the angular velocity $\Omega$; $r$ and $\theta$ are
the radial and colatitude coordinates of point A.}
\label{schema}
\end{figure}

Let us express the rate of change of the angular momentum, ${{\rm d}{\mathcal L}\over{\rm d}t}$, of the element of mass in the volume ABCD represented in Fig.~\ref{schema}:
$${{\rm d}{\mathcal L}\over{\rm d}t}={\bf M},$$
where ${\bf M}$ is the momentum of the forces acting on the volume element. 
We assume that angular momentum is transported only through advection (by a velocity field {\bf U}) and through turbulent diffusion, which may
be different in the radial (vertical) and tangential (horizontal) direction.
The component of the angular momentum aligned with the rotational axis is equal to
\footnote{The components perpendicular to the rotational axis cancel each other when the integration is performed over $\varphi$.} 
$$\underbrace{\rho r^2 \sin\theta {\rm d}\theta {\rm d}\varphi {\rm d} r}_{\rm Mass\ of\ ABCD}\ \ 
\underbrace{r\sin\theta \Omega}_{\rm velocity}\underbrace{r\sin\theta ,}_{\rm distance\ to\  axis}$$ 
where $ \Omega=\dot\varphi$.
Since the mass of the volume element ABCD does not change, the rate of change of the angular momentum can be written
\begin{eqnarray}
\rho r^2 \sin\theta {\rm d}\theta {\rm d}\varphi {\rm d} r {{\rm d}\over {\rm d}t} (r^2 \sin^2\theta \Omega)_{M_r}.
\label{eqn1}
\end{eqnarray}
Due to shear, forces apply on the surfaces of the volume element. The force on the surface AB
is equal to 
$$\underbrace{\eta_v}_{\rm vertical\  viscosity}\ \ \underbrace{r \sin\theta {\partial\Omega\over\partial r}}_{\rm vertical\ shear} 
\underbrace{r^2\sin\theta{\rm d}\theta{\rm d}\varphi}_{\rm surface\  AB}.$$ 
The component of the momentum of this force along the rotational axis is
$$\underbrace{\eta_v r^3\sin^2\theta {\partial\Omega\over\partial r}{\rm d}\theta{\rm d}\varphi}_{\rm force\  on\  AB} 
\ \underbrace{r\sin\theta .}_{\rm distance\  to\  axis}$$
The component along the rotational axis of the resultant momentum of the forces acting on AB and CD is equal to
\begin{eqnarray}
{\partial \over \partial r}(\eta_v r^4 \sin^3\theta {\rm d}\theta{\rm d}{\varphi}{\partial\Omega\over\partial r}){\rm d}r.
\label{eqn2}
\end{eqnarray}
The force on the surface AC due to the tangential shear is equal to
$$\eta_h \underbrace{r\sin\theta {\partial \Omega\over r\partial\theta}}_{\rm tangential\  shear}
\underbrace{r\sin\theta{\rm d}\varphi{\rm d}r }_{\rm surface\  AC},$$
where $\eta_h$ is the horizontal viscosity. The component along the rotational axis of the resultant momentum of the forces acting
on AC and BD is equal to
\begin{eqnarray}
{\partial \over r \partial \theta}(\eta_h r^2\sin^3\theta{\rm d}r{\rm d}\varphi {\partial \Omega \over \partial \theta})r{\rm d}\theta.
\label{eqn3}
\end{eqnarray}
Using Eqs.~\ref{eqn1}, \ref{eqn2} and \ref{eqn3}, simplifying by ${\rm d}r{\rm d}\theta{\rm d}\varphi$, one obtains the equation for the transport of the angular momentum
\begin{eqnarray}
\rho r^2 \sin\theta {{\rm d}\over {\rm d}t} (r^2 \sin^2\theta \Omega)_{M_r}=
{\partial \over \partial r}(\eta_v r^4 \sin^3\theta {\partial\Omega\over\partial r})+
{\partial \over \partial \theta}(\eta_h r^2\sin^3\theta {\partial \Omega \over \partial \theta}).
\label{eqn4}
\end{eqnarray}
Setting $\eta_v=\rho D_v$ and $\eta_h=\rho D_h$ and dividing the left and right member by $r^2\sin\theta$, one obtains
\begin{eqnarray}
\rho{{\rm d}\over {\rm d}t} (r^2 \sin^2\theta \Omega)_{M_r}={\sin^2\theta \over r^2} {\partial \over \partial r}(\rho D_v r^4  {\partial\Omega\over\partial r})+
{1\over \sin\theta}
{\partial \over \partial \theta}(\rho D_h \sin^3\theta {\partial \Omega \over \partial \theta}).
\label{eqn43}
\end{eqnarray}
Now, the left--handside term can be written
$$\rho{{\rm d}\over {\rm d}t} (r^2 \sin^2\theta \Omega)_{M_r}={{\rm d}\over {\rm d}t} (\rho r^2 \sin^2\theta \Omega)_{M_r}-r^2 \sin^2\theta \Omega{{\rm d}\rho\over {\rm
d}t}|_{M_r}.$$
Using the relation between the Lagrangian and Eulerian derivatives, one has
\begin{eqnarray}
\rho{{\rm d}\over {\rm d}t} (r^2 \sin^2\theta \Omega)_{M_r}={\partial\over \partial t} (\rho r^2 \sin^2\theta \Omega)_{r}
+{\bf U}\cdot{\bf \nabla}(\rho r^2 \sin^2\theta \Omega)-r^2 \sin^2\theta \Omega{{\rm d}\rho\over {\rm
d}t}|_{M_r}.
\label{eqn42}
\end{eqnarray}

\noindent Using
$${{\rm d} \rho \over {\rm d}t}|_{M_r}={\partial \rho \over \partial t}|_{r}+{\bf U} \cdot {\bf \nabla} \rho,$$
and the continuity equation
$${\partial \rho \over \partial t}|_{r}=-{\rm div}(\rho {\bf U}),$$
one obtains ${\rm d}\rho/{\rm d}t|_{M_r}+\rho {\rm div}{\bf U}=0$, which incorporated in Eq.~\ref{eqn42} gives
$$\rho{{\rm d}\over {\rm d}t} (r^2 \sin^2\theta \Omega)_{M_r}
={\partial \over \partial t} (\rho r^2 \sin^2\theta \Omega)_{r}+{\bf \nabla}({\bf U}\rho r^2 \sin^2\theta \Omega).$$
Developing the divergence in spherical coordinates and using Eq.~\ref{eqn43}, one finally obtains 
the equation describing the transport of the angular momentum (\citeauthor{Ma98} \citeyear{Ma98}; \citeauthor{Mat04} \citeyear{Mat04} )
\begin{eqnarray}
{\partial\over \partial t} (\rho r^2 \sin^2\theta \Omega)_{r}+{1 \over r^2}{\partial \over \partial r}(\rho r^4\sin^2\theta w_r\Omega)+{1 \over r\sin\theta}
{\partial \over \partial \theta}(\rho r^2\sin^3\theta w_{\theta} \Omega)= \nonumber \\ {\sin^2\theta \over r^2} {\partial \over \partial r}(\rho D_v r^4  {\partial\Omega\over\partial r})+
{1\over \sin\theta}
{\partial \over \partial \theta}(\rho D_h \sin^3\theta {\partial \Omega \over \partial \theta}),
\label{eqn5}
\end{eqnarray}
where $w_r=U_r+\dot r$ is the sum of the radial component of the meridional circulation velocity and the velocity of expansion/contraction, and $w_\theta=U_\theta$,
where $U_\theta$ is the horizontal component of the meridional circulation velocity.
Assuming, as in \citet{Za92}, that the rotation depends little on latitude due to strong horizontal diffusion, we write
$$\Omega(r,\theta)=\bar\Omega(r)+\hat \Omega(r,\theta),$$
with $\hat\Omega \ll \bar\Omega$. The horizontal average $\bar \Omega$ is defined as being the angular velocity of a shell rotating
like a solid body and having the same angular momentum as the considered actual shell. Thus 
$$\bar \Omega={\int \Omega \sin^3 \theta{\rm d} \theta \over \int \sin^3 \theta {\rm d}\theta}.$$
Any vector field whose Laplacian is nul can be decomposed in spherical harmonics. Thus, the meridional circulation velocity can be written
\citep{Mat04}
$${\bf U}=\underbrace{\sum_{l > 0} U_l (r) P_l (\cos\theta)}_{w_r} {\bf e}_r+ \underbrace{
\sum_{l > 0} V_l(r) {{\rm d} P_l (\cos\theta) \over {\rm d} \theta}}_{w_\theta}{\bf e}_\theta,$$
where ${\bf e_r}$ and ${\bf e_\theta}$ are unit vectors along the radial and colatitude directions respectively.
Multiplying Eq.~\ref{eqn5} by $\sin\theta{\rm d}\theta$ and integrating it over $\theta$ from 0 to $\pi$, one obtains
\citep{Ma98}
\begin{eqnarray}
{\partial \over \partial t}(\rho r^2 \bar \Omega)_r={1 \over 5 r^2}{\partial \over \partial r}(\rho r^4 \bar \Omega [U_2(r)-5\dot r])
+{1 \over r^2}{\partial \over \partial r}\left(\rho D_v r^4 {\partial \bar \Omega \over \partial r} \right).
\label{eqn6}
\end{eqnarray}
It is interesting to note that only the $l=2$ component of 
the circulation is able to advect a net amount of angular momentum. As explained in
\citet{Spie92} the higher order components do not contribute to the vertical transport of angular momentum.
Note also that the change in radius $\dot r$ of the given mass shell is included in Eq.~\ref{eqn6}, which is the Eulerian formulation of the
angular momentum transport equation. In its Lagrangian formulation, the variable $r$ is linked to $M_r$ through ${\rm d}M_r=4\pi r^2 \rho {\rm d}r$, and
the equation for the transport of the angular momentum can be written
\begin{eqnarray}
\rho{\partial \over \partial t}(r^2 \bar \Omega)_{M_r}={1 \over 5 r^2}{\partial \over \partial r}(\rho r^4 \bar \Omega U_2(r))
+{1 \over r^2}{\partial \over \partial r}\left(\rho D_v r^4 {\partial \bar \Omega \over \partial r} \right).
\label{eqn7}
\end{eqnarray}
The characteristic time associated to the transport of $\Omega$ by the circulation is \citep{Za92}
\begin{eqnarray}
t_\Omega\approx t_{KH} \left({\Omega^2 R \over g_s}\right)^{-1},
\label{eqn8}
\end{eqnarray}
where $g_s$ is the gravity at the surface and $t_{KH}$ the Kelvin--Helmholtz timescale, which is the characteristic timescale for the change
of $r$ in hydrostatic models. From Eq.~\ref{eqn8}, one sees that $t_\Omega$ is a few times $t_{KH}$, which itself is much shorter that
the Main Sequence lifetime. 

In order to resolve Eq.~\ref{eqn7}, one needs expressions for  $U_2 (r)$, $D_v$, and $D_h$. 
In the following we indicate the general lines of the reasoning and give 
the references where more complete mathematical derivations can be found.
The velocity of meridional circulation is derived from the equation of energy conservation \citep{Mes53}
\begin{eqnarray}
\rho T\left[{\partial S \over \partial t}+( \dot r {\bf e}_r+{\bf U})\cdot{\bf \nabla}S\right]={\rm div}(\chi{\bf \nabla}T)+\rho\epsilon-{\rm div}{\bf F}_h
\label{eqn9}
\end{eqnarray}
where $S$ is the entropy per unit mass,
$\chi$ the thermal conductivity, $\epsilon$ the rate of nuclear energy per unit mass and ${\bf F}_h$ the flux of thermal energy due to
horizontal turbulence. All the quantities are expanded linearly around their average on a level surface or isobar, using
Legendre function of order 2 $P_2(\cos\theta)$.
For instance
$$T(P,\theta)=\bar T(P)+\tilde{T}P_2(\cos\theta).$$
Then  Eq.~\ref{eqn9} is linearized and an expression
for $U_2$ can be deduced \citep{Za92}. Using the same method 
\citet{Ma98} revised the expression for $U_2$ to account for expansion and contraction in non--stationary models. They also studied the effects of the 
$\mu$--gradients (mean molecular weight gradients), of the horizontal turbulence and considered a general equation of state. They
obtained 
\begin{eqnarray}
U_2(r)={P \over \bar \rho \bar g
C_P\bar T[\nabla_{\rm ad}-\nabla+(\varphi/\delta)\nabla_\mu ]}
\times\left[{L\over M_*}(E_\Omega+E_\mu)+{C_P \over \delta}{\partial \Theta \over \partial
t}\right],   
\label{eqn10}
\end{eqnarray}
where $M_{\star}=M \left( 1 - \frac{\Omega^2}{2 \pi G \rho_{\rm{m}}}  \right)$
is the reduced mass and the other symbols have the same meaning as in \citet{Za92} and \citet{Ma98}.
The driving term in the square brackets in the second member
 is $E_{\Omega}$.  It behaves mainly like
$E_{\Omega}  \simeq  \frac{8}{3} \left[ 1 - \frac{{\Omega^2}}
{2\pi G\overline{\rho}}\right] \left( \frac{\Omega^2r^3}{GM}\right)$.
The term $\overline{\rho}$ means the average
on the considered equipotential.
The term with the minus sign in the square bracket is the 
Gratton--\"{O}pik term, which becomes important in the outer layers
when the local density is small. This term  produces negative values of $U_2(r)$
(noted $U(r)$ from now), meaning that 
the circulation is going down along the polar axis and up 
in the equatorial plane. This makes an outward transport 
of angular momentum, while a positive $U(r)$ gives an inward transport.
At lower $Z$, the Gratton--\"{O}pik term is negligible, which contributes
to make larger $\Omega$--gradients in lower $Z$ stars. 

Recently \citet{Mat04} 
rederived the system of partial differential equations, which govern the transport
of angular momentum, heat and chemical elements. They 
expand the departure from spherical symmetry to higher order and include explicitly the differential rotation 
in latitude, to first order. Boundary conditions for the surface and at 
the frontiers between radiative and convective zones are
also explicitly given in this paper.

In the above equation $D_v$ is the shear diffusion coefficient whose expression is taken as in \citet{Ta97}.
The usual estimate of $D_{\mathrm{h}} =
\frac{1}{c_{\mathrm{h}}} r \;|2V(r) - \alpha U(r)| $ was given by 
\citet{Za92}. Recent studies suggest that this coefficient 
is at least an order of magnitude larger (\citeauthor{Ma03} \citeyear{Ma03}; 
\citeauthor{Mati04} \citeyear{Mati04}).

\section{Equations for the transport of the chemical species}

The derivation of the equation for the transport of chemical species
proceeds much like that of angular momentum: the mass fraction $X_i$ of a given element
$i$ obeys an advection--diffusion equation which can be written
\begin{eqnarray}
{\partial (\rho X_i) \over \partial t}+{\rm div}(\rho X_i {\bf U})=\ \ \ \ \ \ \ \ \ \            \hfill\hfill \nonumber \\
{1 \over r^2}{\partial \over \partial r}\left(\rho D_v r^2 {\partial X_i \over \partial r}\right)
+{1 \over r^2 \sin\theta}{\partial \over \partial \theta}\left(\sin\theta \rho D_h {\partial X_i \over \partial \theta}\right).
\label{eqn11}
\end{eqnarray}
Splitting the mass fractions in their horizontal average ($\bar X_i$) and their fluctuation on the isobar, assuming that
$D_h \gg D_v$, \citet{Ch92} showed that Eq.~\ref{eqn11} can be written
\begin{eqnarray}
\rho\left(
{{\rm d} \bar X_i \over {\rm d} t} 
\right)_{M_r}=
{1 \over r^2}
{\partial \over \partial r}
\left(\rho(D_v+D_{\rm eff})r^2 {\partial \bar X_i \over \partial r} \right)
\label{eqn12}
\end{eqnarray}
with the effective diffusivity

$$D_{\rm eff} = \frac{\mid rU(r) \mid^2}{30 D_h}.$$ 
$D_{\rm eff}$ expresses the 
effects of the meridional circulation and of the strong horizontal turbulence. 
This expression of $D_{\rm eff}$ tells us that the vertical
advection of chemical elements is  inhibited by the
strong horizontal turbulence characterized by $D_{\rm{h}}$. 
For simplicity in Eq.~\ref{eqn12}, the terms due to nuclear reactions are omitted, as well as the effects
of element separation through radiation and gravitational settling.
Let us finally add that various numerical procedures for resolving the diffusion equation
are described and compared in \citet{Meyn04}.

%$$A=-H_T {{\rm d} \over {\rm d}r}\left({\Theta \over \delta}-{\varphi \over \delta}\Lambda\right)
 %   +{\Theta \over \delta}-\left(\chi_\mu +{\varphi \over \delta}\chi_T +{\varphi \over \delta}  \right)\Lambda$$
 
\section{Effects of rotation on the stellar winds}

The taking into account of the von Zeipel theorem in the frame of  the radiative driven wind theory 
has many interesting consequences (\citeauthor{Ow96} \citeyear{Ow96}; 
\citeauthor{Ma99} \citeyear{Ma99}; \citeauthor{MMVI} \citeyear{MMVI}):
\begin{enumerate}
\item The maximum luminosity of a rotating star is reduced by rotation.
\item The expression for the critical velocity in a rotating star is different
if the star is near or far from the Eddington limit. 
Far from the Eddington limit,  $v_{\mathrm{crit,1}} = 
\left( \frac{2}{3} \frac{GM}{R_{\mathrm{pb}}} \right)^{\frac{1}{2}}$, where
$R_{\mathrm{pb}}$ is the polar radius at break--up. When
the Eddington factor is bigger than 0.639, the expression for $v_{\mathrm{crit}}$
is different \citep{MMVI}. It is
equal to 0.85, 0.69, 0.48, 0.35, 0.22, 0 times $v_{\mathrm{crit, 1}}$ for $\Gamma=$
0.70, 0.80, 0.90, 0.95, 0.98 and 1.00 respectively.
\item Rotation induces anisotropies of the stellar winds
(\citeauthor{MD01} \citeyear{MD01}; \citeauthor{Dw02} \citeyear{Dw02}).
\item Rotation increases the mass loss rates. We may estimate the mass loss rates of a rotating star 
compared to that of a non--rotating star at the same location in the HR
diagram. The result is \citep{MMVI}
\begin{equation}
\frac{\dot{M} (\Omega)} {\dot{M} (0)} \simeq
\frac{\left( 1  -\Gamma\right)
^{\frac{1}{\alpha} - 1}}
{\left[ 1 - 
\frac{4}{9} (\frac{v}{v_{\mathrm{crit, 1}}})^2-\Gamma \right]
^{\frac{1}{\alpha} - 1}} \; ,
\end{equation}
\noindent
where $\Gamma$ is the electron scattering opacity for a non--rotating
star with the same mass and luminosity, $\alpha$ is a force multiplier \citep{Lamers95}. 
For B--stars far from the Eddington Limit, $\frac{\dot{M} (\Omega)} {\dot{M} (0)} \simeq 1.5$
For stars close to
$\Gamma=1$ the  increase of the mass loss rates may reach orders of magnitude. 
\end{enumerate}
It is interesting to note that three kinds of break--up limit can be defined depending on which
mechanism, radiative acceleration, centrifugal acceleration or both contribute to counterbalance
the gravity: 1.-- The
$\Gamma$--Limit occurs when radiation effects largely dominate; 2.-- The $\Omega$--Limit, when
rotation effects are essentially determining break--up; 3.-- The $\Omega \Gamma$--Limit, when both
rotation and radiation are important for the critical velocity. In the cases of Luminous blue variables
the $\Gamma$-- and $\Omega \Gamma$--Limits likely occurs, while in the case of Be stars,
the $\Omega$--Limit is the relevant limit.

\section{Some successes of rotating models}

The inclusion of rotation in massive star models improves in many respects
the theoretical predictions. In particular, rotating models can reproduce
the chemical enrichments observed at the surface of OBA stars (\citeauthor{Hel00} \citeyear{Hel00}; \citeauthor{MMV} \citeyear{MMV} ), 
the number ratio of blue to red supergiants in the Small Magellanic Cloud
\citep{MMVII}, the variation with the metallicity of the number ratio of Wolf--Rayet to O type stars (\citeauthor{MMX} \citeyear{MMX}) as well
as of the type Ibc to type II supernovae (\citeauthor{Pr03} \citeyear{Pr03}; \citeauthor{MMXI} \citeyear{MMXI}). 
Rotating models predict new chemical yields (\citeauthor{HMM4} \citeyear{HMM4}) and
naturally lead to the production
of primary nitrogen at low metallicity \citep{MMVIII}, they also bring new views on the evolution of pop III stars
\citep{MME04}.
Comparisons with the observed rotational velocities of stars become possible and interesting
consequences for the rotation rate of young pulsars and the progenitors of Gamma Ray Burst can
be obtained (\citeauthor{WH04} \citeyear{WH04}; \citeauthor{MMH04} \citeyear{MMH04}).

\section{Conditions for reaching break--up during the MS phase}

Recently \citet{Town04} have given strong arguments indicating that the $\upsilon \sin i$
measured for Be stars might be systematically underestimated. An important implication of their work
is that these stars may be rotating
much closer the critical velocity than generally assumed.
As the authors note, rotation close to critical
effectively reduces the effective equatorial gravity and could make 
material to easily leak into a disc. Typically, when $\upsilon/\upsilon_{\rm crit} \sim 0.95$, the velocity necessary to launch material into
orbit is of the same order as the speed of sound in the outer layers. Therefore
mechanisms, as nonradial pulsation,
begin to be effective for orbital ejection \citep{Owocki}.
In view of these considerations, it appears interesting to investigate with our rotating stellar models, which are
the initial conditions required for a star to reach the critical velocity during the MS phase. This is the object
of this section.

\begin{figure}[!t]
\plottwo{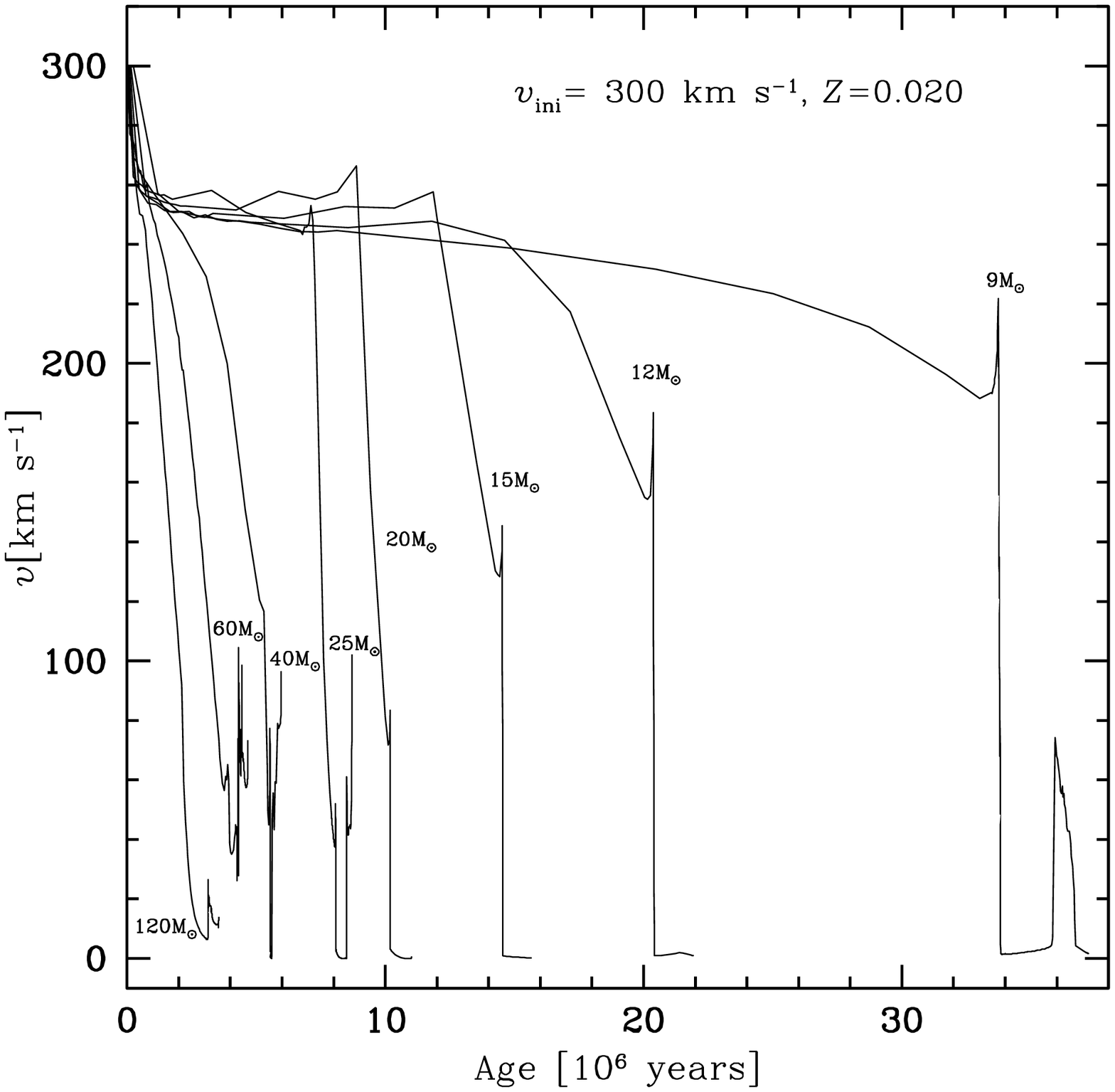}{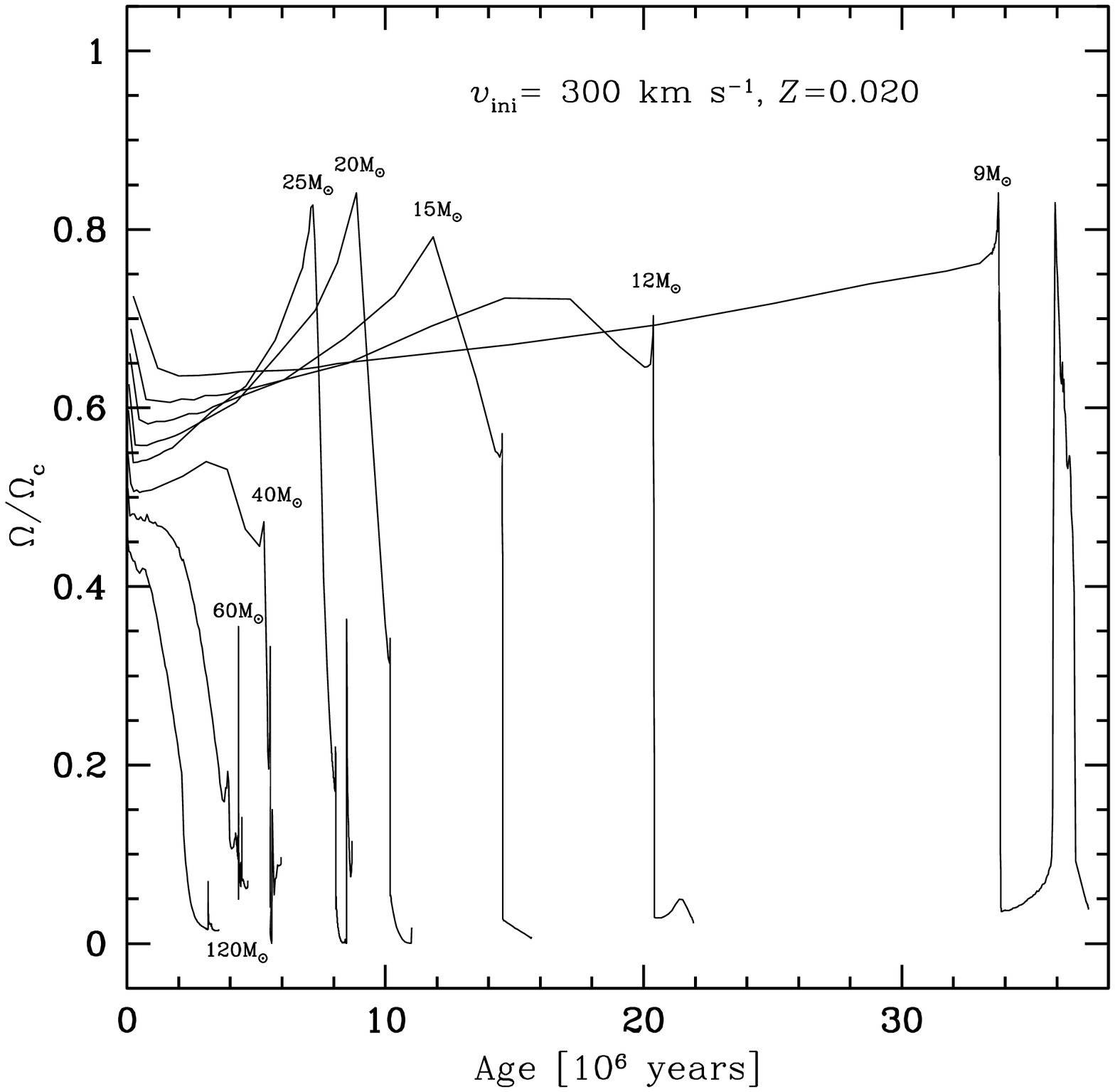}
\caption{{\it Left :}Evolution of the rotational velocities for star models 
  of different initial masses between 120 and 9 M$_{\odot}$ with account taken
  of anisotropic mass loss during the MS phase. The rapid decrease near the end 
  of the MS evolution is due to the bi--stability limit in the mass loss rates.
   The location  of the end of the MS phase corresponds to the vertical decrease of the velocity. {\it Right :}
   Evolution of the fraction 
  $\frac{\Omega}{\Omega_{\mathrm{c}}}$ of the angular velocity to 
  the critical angular velocity at the surface   of  star models 
  of different initial masses between 120 and 9 M$_{\odot}$ with account 
  taken of anisotropic mass loss during the MS phase.}
\label{vsur}
\end{figure}

Figs.~\ref{vsur} show the evolution of the rotational velocities and of the fraction 
 $\frac{\Omega}{\Omega_{\mathrm{c}}}$ of the angular velocity to 
 the critical angular velocity at the surface   of  star models 
 of different initial masses between 9 and 120 M$_{\odot}$ with account taken
 of anisotropic mass loss during the MS phase.
The evolution of the rotation velocities at the stellar surface depends mainly 
on 2 factors, the internal coupling and the mass loss.
 
1.-- The  coupling mechanisms  transport angular momentum
in the stellar interiors. The extreme case of strong coupling is
the classical case of solid body rotation. In this case when mass loss is small, the star  
reaches the critical velocity during the MS phase more or less quickly depending on the
initial rotation as shown by \citet{Sa70} (see also \citeauthor{La97} \citeyear{La97}).
In the case of no coupling, i.e. of local conservation of the angular momentum,
rotation becomes more and more subcritical. In the present models, with initial velocities at the beginning
of the MS phase of $\sim$ 260 km s$^{-1}$ the
situation  is  intermediate, with a moderate coupling 
due mainly to meridional circulation, which is more efficient 
\citep{MMV} than shear transport,
as far as transport of angular momentum is concerned.  

\begin{figure}[!t]
\plotfiddle{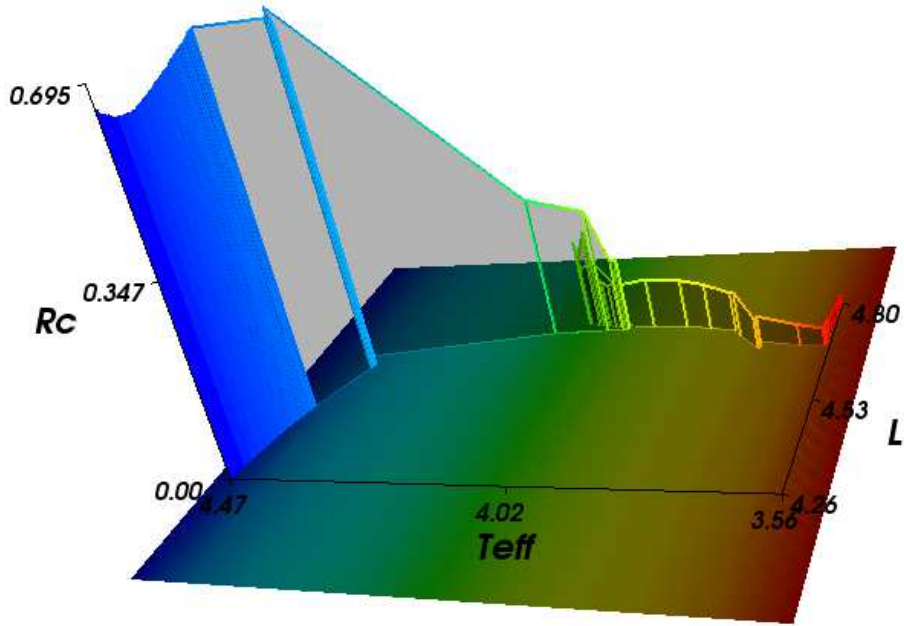}{9cm}{0}{60}{60}{-190}{-120}
%\plotone{B15rc.ps}
\caption{Three dimensional evolutionary track of a 15 M$_\odot$ model at solar metallicity, with
a velocity at the beginning of the MS phase of 260 km s$^{-1}$. The evolution in the theoretical HR diagram
is plotted on the horizontal plane. The vertical axis shows the evolution of the surface angular velocity devided
by the critical or break--up angular velocity ($R_c=\Omega/\Omega_{\rm crit}$).}
\label{omomc}
\end{figure}

2.-- For a given degree of coupling, {\emph{the mass loss rates play a most critical role 
in the evolution of the surface rotation}}.
As shown by the comparison of the
models at $Z=0.02$ and $Z=0.004$ \citep{MMVII}, for masses greater
than $\sim$40 M$_{\odot}$ the models with solar composition have velocities that decrease
rather rapidly, while at $Z=0.004$ the velocities go up. Thus, for the most massive stars 
with moderate or strong coupling, the mass loss rates are the main factor influencing 
the evolution of rotation. The effect of mass loss
is also well illustrated in Fig.~\ref{omomc}, which shows the evolution of $\Omega / \Omega_{\rm crit}$
at the surface of a 15 M$_\odot$ solar metallicity model. One sees that during the MS phase, the
ratio increases as a function of time, until $\log T_{\mathrm{eff}}$ becomes inferior to 4.40.
According to Vink et al. (\citeyear{Vink00}; \citeyear{Vink01}),
below this value there is a sudden increase of the mass loss rates, which makes
the rotation velocities rapidly decrease.

It is interesting to note that 
below a mass of about 12 M$_{\odot}$, the mass loss rates are smaller and the internal 
coupling plays the main role in the evolution of the rotational velocities. This
provides an interesting possibility of tests on the internal coupling by studying
the differences in rotational velocities for stars at different distances of the 
ZAMS. In particular, such a study could allow us  to test the role
of magnetic coupling in radiative envelopes, which is now a major open question
in stellar rotation studies \citep{spruit02}.

\begin{figure}[!t]
\plottwo{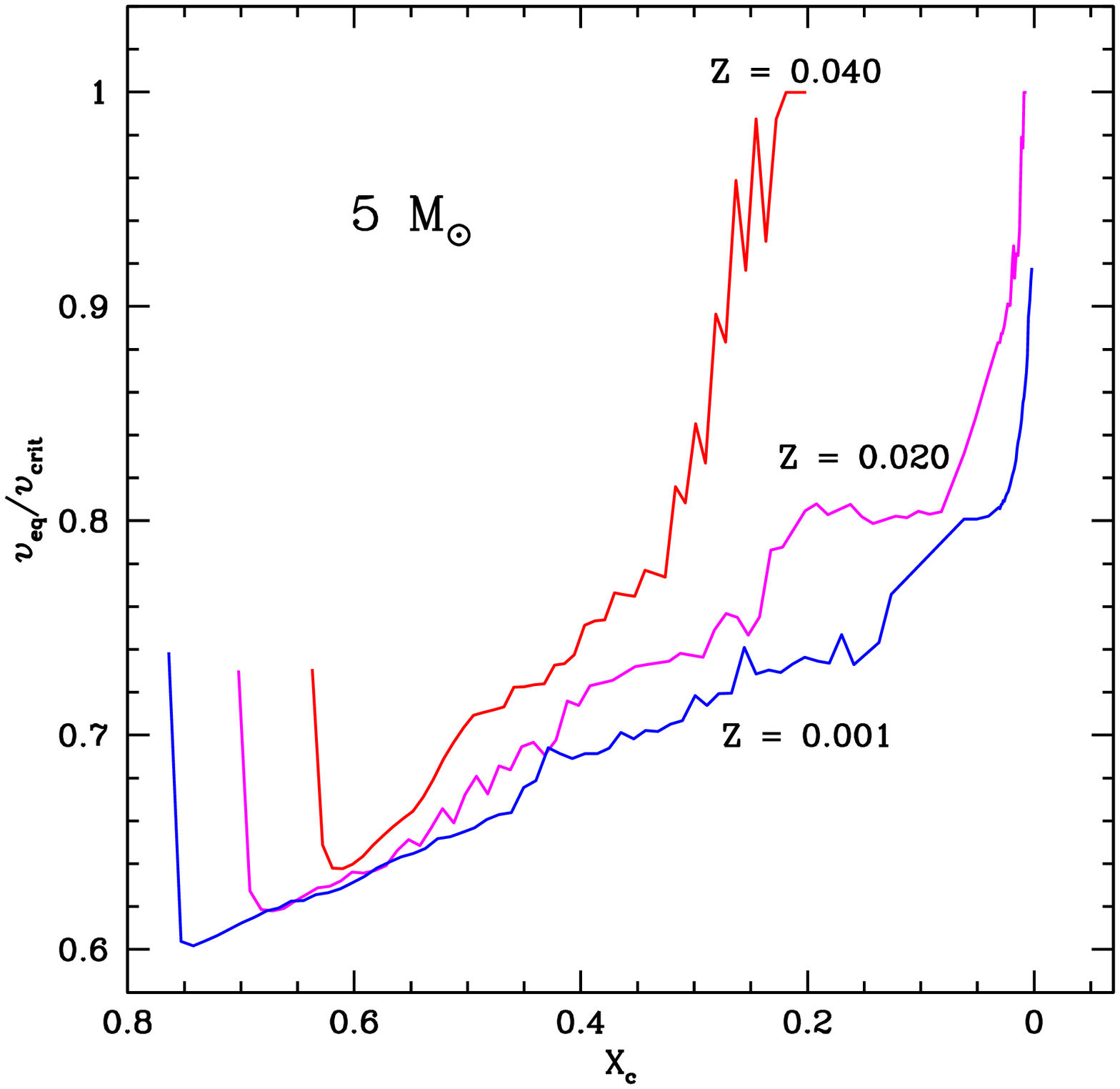}{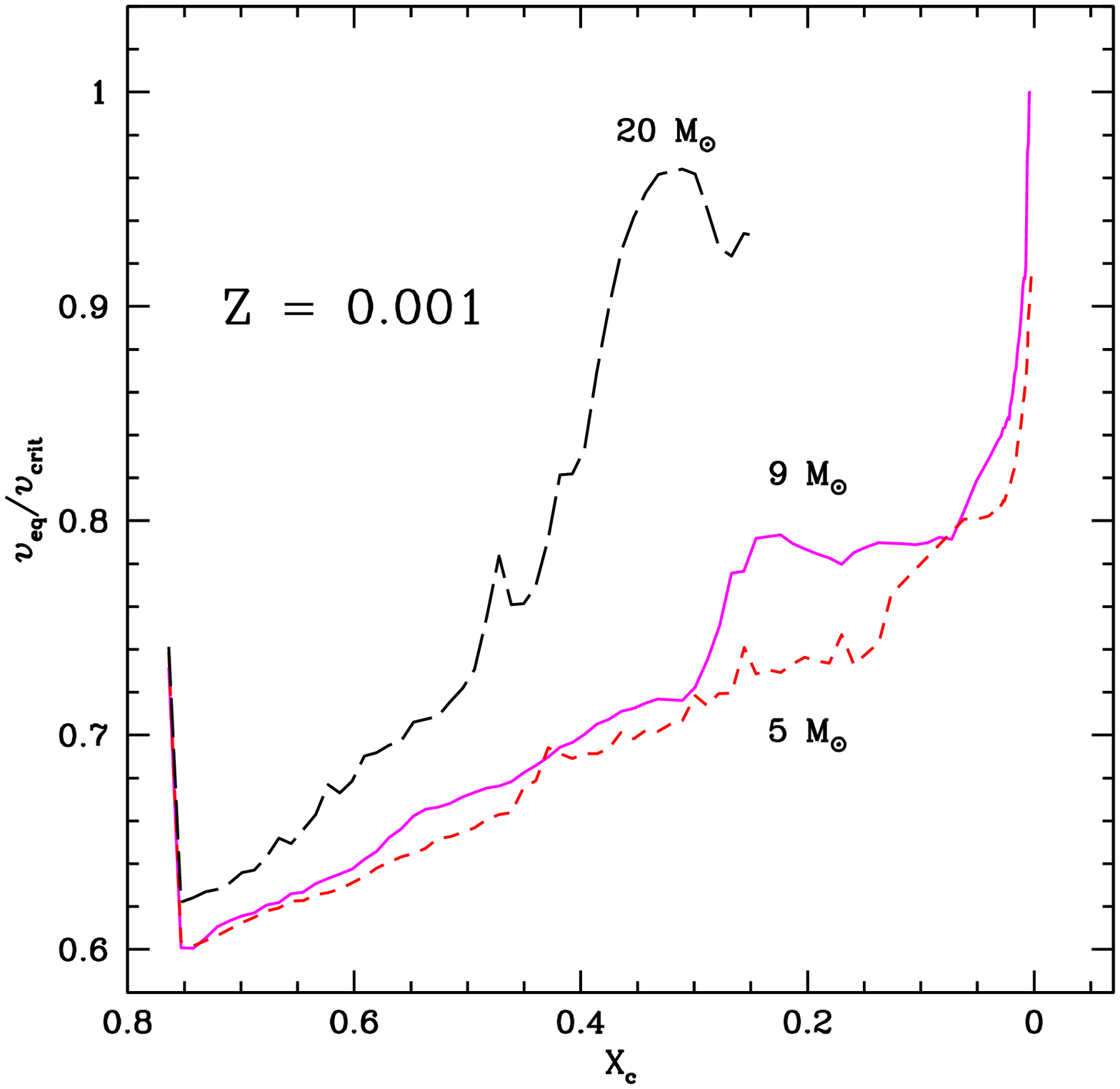}
\caption{{\it Left :} Evolution of the equatorial velocity divided by the critical velocity at the
surface of 5 M$_\odot$ stellar models at different metallicities. {\it Right\~:} Same as in the left panel, for
different initial mass stellar models at the metallicity Z = 0.001. The initial values of
$\upsilon_{\rm eq}/\upsilon_{\rm crit}$ are between 0.60 and 0.64.}
\label{crit}
\end{figure}

From Fig.~\ref{vsur}, we see that the stars which have a velocity around 260 km s$^{-1}$ at the beginning of the MS phase, do not reach
the break--up limit.
Taking into account of the wind anisotropies induced by rotation, 
a 9 M$_\odot$ stellar model at solar metallicity with an equatorial velocity
at the beginning of the core H--burning phase equal to 340 km s$^{-1}$
reaches the break--up limit during
the MS phase. This corresponds to
values of $\Omega/\Omega_{\rm crit}$  and of $\upsilon/\upsilon_{\rm crit}$
equal to 0.81 and 0.61 respectively. The model with with an initial equatorial velocity
of 290 km s$^{-1}$ ($\Omega/\Omega_{\rm crit}=0.71$, $\upsilon/\upsilon_{\rm crit}=0.52$),
approaches but does not reach the break--up limit during the MS phase.
At the end of the MS phase, the 290 and 340 km s$^{-1}$ models
present N/C ratios at the surface equal to 2.8 and 3.2 times the initial value.

In Fig.~\ref{crit}, evolutions of  $\upsilon/\upsilon_{\rm crit}$ for different initial metallicities and  masses
are shown. Interestingly, one notes that the 5 M$_\odot$ stellar model at Z = 0.040 reaches the break--up limit well before the
model at Z=0.020. This comes essentially from the dependence on density of the meridional velocity as given
by the Gratton-\"Opik term. At higher metallicity, the radii of stars are bigger and thus
the density in the outer layers are smaller.
At a given metallicity (see right panel of Fig.~\ref{crit}), 
the higher initial mass star reach the break--up limit before
the lower initial mass models. Again this can be 
explained through the density dependance of the meridional circulation velocity.

Thus these numerical experiments show the following:

1) For stars suffering little mass loss, the main effect which can bring these stars near the break-up limit is the 
outwards transport of angular momentum by the meridional circulation. Since the velocity of the meridional currents
in the outer layers scales with the inverse of the density, the process becomes more efficient
for stars of higher initial mass and/or higher initial metallicity.

2) When the metallicity increases however, mass loss becomes more and more important and can prevent stars
to reach the break--up limit.

How the fraction of stars which reach the break--up limit varies as a function of the initial metallicity
requires a good knowledge of the distribution of the initial velocities as a function of the mass and of the metallicity.
Large and extended surveys are being now performed with this purpose
(see e.g.  \citeauthor{No04} \citeyear{No04}; \citeauthor{Ro04} \citeyear{Ro04}). 
Also the study of the time evolution of the surface velocities on the Main Sequence for stars
undergoing little mass loss would provide very important hints on the efficiency of the
internal tranport mechanisms of the angular momentum. 
\begin{acknowledgements}
 We express our thanks to Raphael Hirschi for his help in producing
 Figure~\ref{omomc}.
\end{acknowledgements}

\end{document}